\documentstyle[aps,preprint,epsf]{revtex}
\begin{document} 
\preprint{WIS98/22/Aug. -- DPP (quant-ph/9808058)}
\draft
\title{Dephasing and collapse in continuous measurement of a single system}
\author{S.A. Gurvitz}

\address{Department of Particle Physics, Weizmann Institute of
       Science, Rehovot 76100, Israel}
\date{\today}
\maketitle
\begin{abstract}  
We show that long standing debates on the collapse and 
the role of the observer in quantum mechanics can 
be resolved experimentally via a nondistructive 
continuous monitoring of a single quantum system.
An example of such a system, coupled with 
the point-contact detector is presented. 
The detailed quantum mechanical analysis of the entire 
system (including the detector) shows that 
under certain conditions the measurement collapse 
would generate distinctive effects in the detector 
behavior, which can be experimentally investigated. 
\end{abstract}
\vskip 1cm
\section{Introduction}
According to the principles of quantum mechanics, 
a system in the linear superposition of different states collapses 
to one of the states after the measurement. This is the
wave-function collapse\cite{neu}, which 
has been debated since the early days of quantum mechanics. 
The main question is of whether the collapse is originated 
by the interaction with detector. More precisely, is   
it described by  the Schr\"odinger equation,   
$i\dot\rho =[{\cal H},\rho ]$, applied to the entire system. Here
$\rho ({\cal S},{\cal S}';{\cal D},{\cal D}',t)$ is the total 
density-matrix, where ${\cal S}({\cal S}')$ and ${\cal D}({\cal D}')$ 
are the variables of the measured system and the detector respectively, and  
${\cal H}$ is the total Hamiltonian. 
  
In order to determine of how the detector affects 
the measured system one needs to
``trace out'' the detector variables in the total density matrix,
\begin{equation}
\sum_D\rho ({\cal S},{\cal S}',{\cal D},{\cal D},t)\to
\sigma ({\cal S},{\cal S}',t).
\label{in1}
\end{equation}
Since the detector is a macroscopic system, its density of states 
is very high (continuum). In this case the tracing generates 
an exponential damping of the 
off-diagonal terms (${\cal S}\not ={\cal S}'$) (``decoherence'').
As a result the reduced density-matrix of the 
observed system becomes the statistical mixture during the measurement,   
$\sigma ({\cal S},{\cal S}',t)\to\bar\sigma ({\cal S},{\cal S}',t)
\delta_{{\cal S},{\cal S}'}$. 
The latter tells us that the system is actually in 
one of its states with the corresponding probability 
$\bar\sigma({\cal S},{\cal S},t)$. Notice that  
such a tracing can be performed at any time $t$ with no distortion of  
the Schr\"dinger equation of motion for the {\em entire} system, 
$i\dot\rho =[{\cal H},\rho ]$ \cite{gp,gur1}. Therefore the 
unitary evolution of the {\em entire} system is not violated. 

Intensive investigations during last years demonstrated that 
in many cases the collapse can be attributed to  
the decoherence only\cite{zur}. 
Nevertheless, the real  problem appears in a description of  
continuous nondistructive measurement of a {\em single} quantum object.
Consider, for example, an electron oscillating between
two different states ($a$ and $b$) that are continuously monitored. 
We assume that these states are correlated with the macroscopically 
distinctive states $A$ and $B$ of the detector. As a result of  
interaction with the macroscopic detector,  
the electron oscillations are damped, so its reduced 
density-matrix approaches the statistical mixture, 
Eq.~(\ref{in1}). In this limit the detector displays 
one of the states $A$ or $B$. It tells us that the electron 
is found in one of the states, respectively, $a$ or $b$. 
Then due to the measurement collapse the electron reduced density-matrix 
becomes a pure state $\sigma ({\cal S},{\cal S}',t)\to
\delta_{{\cal S},a}\delta_{{\cal S}',a}$ (the detector displays the 
state $A$). In this case the electron starts to oscillate again 
until the next quantum jump takes a place\cite{knight} etc.  
  
It is rather clear that these quantum jumps  
cannot be attributed to the decoherence only. Moreover,
their appearance is directly related to the 
wave function collapse. This makes the study of
a single quantum system under constant monitoring 
especially important
for understanding of the measurement problems\cite{gur2}.  
In addition, such a study have even practical applications. 
This is in view of a possible use of single quantum systems for quantum 
computing\cite{sasha1}. 

An essential point, which is missed in many studies 
of quantum measurements is a detailed quantum mechanical treatment 
of the entire system, that is, of the detector and the measured 
system together\cite{ander}. 
The reason is that the detector is a macroscopic device, the quantum 
mechanical analysis of which is rather complicated. 
Thus one can expect that the mesoscopic systems,
which are between the microscopic and 
macroscopic scales, would be very useful for 
this type of investigation\cite{imry}.
A generic example of such a system 
has been considered in\cite{gur1}. 
It consisted of two coupled quantum dots,
occupied by one electron, and the point-contact detector\cite{pep}, 
monitoring the occupation of one of the dots\cite{buks}. 
The system has been analyzed by using the Bloch-type rate 
equations for the density-matrix, obtained directly from the 
many-body Schr\"odinger equation\cite{gp,gur1}.
These equations describe the behavior of both,  
the observed electron and the macroscopic (mesoscopic) detector,  
in the most transparent and simple way. 

In this paper we perform such a ``simultaneous'' microscopic study 
of the detector and the measured system during the continuous 
measurement for the setup proposed in\cite{gur1,gur2}. 
This analysis would explicitly show where the problem of 
the wave function collapse emerges and in what way it affects
the experimental outcome. Therefore it may open a possibility 
for experimental investigation of the measurement problem. 
 
The plan of the paper is as follows: In Sect. 2 we describe the 
Bloch-type rate equation for the entire system and their 
quantum-mechanical microscopic origin. 
In Sect. 3 we discusses the microscopic behavior 
of the detector, separated from the measured system.  
In Sect. 4 we concentrate on the quantum mechanical behavior 
of the detector during the measurement process. 
The necessity for the measurement collapse, is explained. 
We discuss its different scenarios and their experimental consequences. 
The last section is Discussion. The details of quantum mechanical derivation 
of the classical rate equations for the point-contact detector 
are given in Appendix, as well as the evaluation of the 
average current and the current fluctuation. 

\section{Continuous monitoring of a single electron with the point-contact 
detector}

\subsection{General description}

Consider the measurement of a single electron oscillating in the double-dot 
by using the point-contact detector\cite{pep,buks}. 
Such a set up is shown schematically in Fig.~1, where
 
\vskip.4cm
\begin{minipage}{13cm}
\begin{center}
\leavevmode
\epsfxsize=13cm
\epsffile{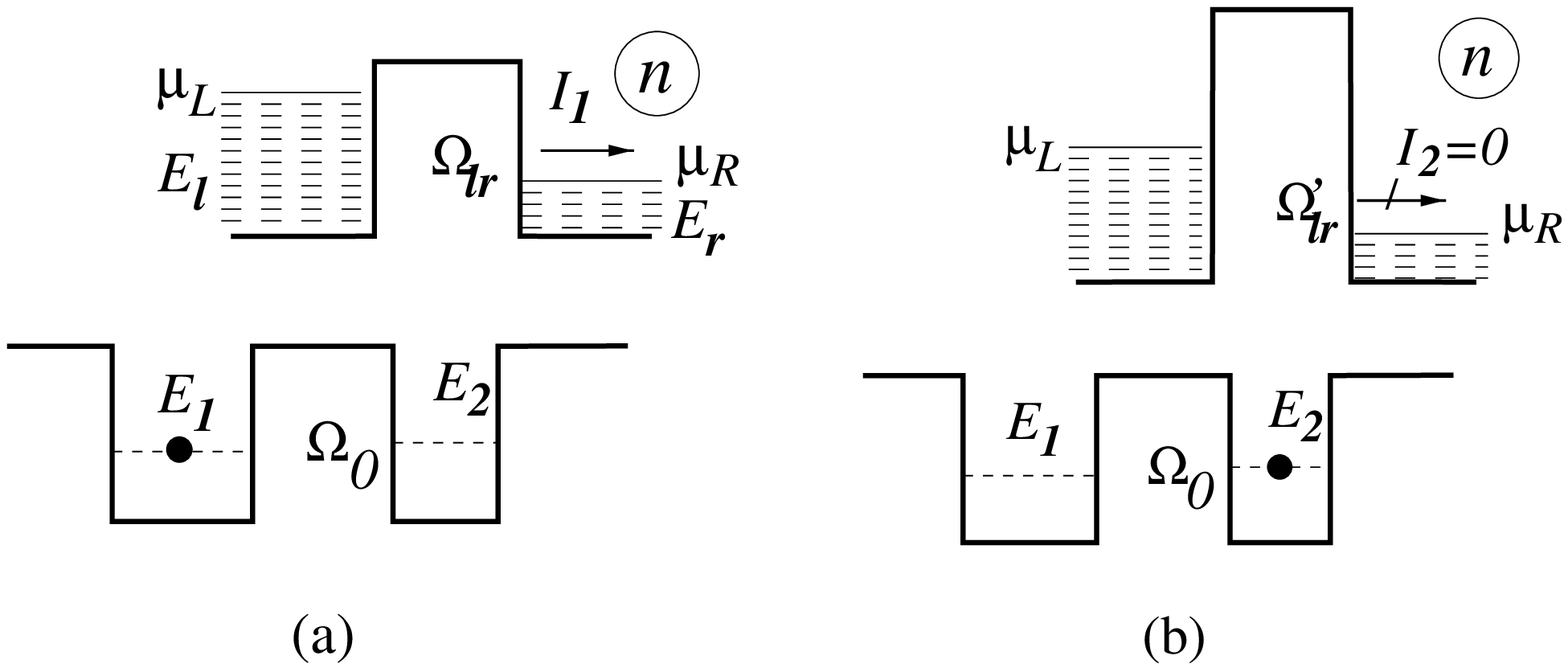}
\end{center}
{\begin{small}
Fig.~1. The point-contact detector near the double-dot.
$\Omega_{lr}$ is the coupling between the level $E_l$ and 
$E_r$ in the left and the right 
reservoirs. $\Omega_0$ is the coupling between the quantum dots.
The index $n$ denotes the number of electrons penetrating 
to the right reservoir (collector) at time $t$.
\end{small}}
\end{minipage} \\ \\ \\
the point-contact, represented by the barrier, is 
placed near one of the dots. 
The barrier is connected with two 
reservoirs at the chemical potentials 
$\mu_L$ and $\mu_R=\mu_L-V_d$ respectively, where 
$V_d$ is the applied voltage. 
Since $\mu_L > \mu_R$, the current $I=eTV_d/(2\pi )$ 
flows through the point-contact\cite{land}, 
where $e$ is the electron charge and $T$ is the transmission coefficient 
of the point-contact.  (We choose the units where $\hbar =1$).
The penetrability of the point-contact (the barrier height)
is modulated by the electron, 
oscillating inside the double-dot.  
When the electron occupies the left dot, the transmission coefficient is 
$T_1$. However, when the 
right dot is occupied, the transmission coefficient 
$T_2\ll T_1$ due to the electrostatic repulsion generated by the electron. 
As a result, the current $I_2\ll I_1$. Without loosing generality we 
assume that $T_2=0$, so that the point contact is blocked whenever the 
right dot is occupied. 
Since the difference $\Delta I=I_1-I_2$
is macroscopically large, one can determine which of the dots 
is occupied by observing the point-contact current. 
Yet, the entire system can be treated quantum-mechanically. 
For its description we use the following tunneling 
Hamiltonian\cite{gur1}:
\begin{equation}
{\cal H}={\cal H}_{PC}+{\cal H}_{DD}+{\cal H}_{int},
\label{a1}
\end{equation}
where 
\begin{mathletters}
\label{a2}
\begin{eqnarray}
{\cal H}_{PC}&=&\sum_l E_la_l^\dagger a_l+\sum_r E_ra_r^\dagger a_r
+\sum_{l,r}\Omega_{lr}(a_l^\dagger a_r +a_r^\dagger a_l )\, , 
\label{a2a}\\
{\cal H}_{DD} &=& E_1 c_1^{\dagger}c_{1}+E_2 c_2^{\dagger}c_{2}+
              \Omega_0 (c_2^{\dagger}c_{1}+ c_1^{\dagger}c_{2})\, ,
\label{a2b}\\
{\cal H}_{int}&=&-\sum_{l,r}\Omega_{lr}c_2^{\dagger} 
c_2(a^{\dagger}_la_r +a^{\dagger}_ra_l)\, .
\label{a2c}
\end{eqnarray}
\end{mathletters}
Here ${\cal H}_{PC}$, ${\cal H}_{DD}$ and ${\cal H}_{int}$ are the 
Hamiltonians describing the point-contact, double-dot and their 
mutual interaction, respectively, $E_{l,r}$ are the energy levels 
in the left (right) reservoir, and $\Omega_{lr}$ is the coupling between 
the reservoirs. It is related to the penetration coefficient by
$(2\pi )^2\Omega^2\rho_L\rho_R=T$, where $\rho_{L,R}$ 
are the density of states in the left (right) reservoir\cite{bard}. 
If the left well is occupied, the coupling   
$\Omega'_{lr}=0$ 
due to the interaction term, ${\cal H}_{int}$. 
In fact, the Hamiltonian ${\cal H}$ should include additional terms
for a macroscopic device that actually counts the charge transmitted to 
the detector (the pointer). This question is considered later in a more 
detail. In any case such a pointer would not affect the (macroscopic) 
detector current\cite{levi}, and therefore the observed electron. 
It implies that we can consider the 
measurement setup shown in Fig.~1 as the {\em closed} system,
describing by the Hamiltonian (\ref{a1}). 

For simplicity we consider the reservoirs at zero temperature 
and the entire system in a 
pure state, i.e. we describe it by the wave-function.  
The latter can be written as
\begin{eqnarray}
|\Psi (t)\rangle &=& \exp (-i{\cal H}t)|0\rangle =
\left [ b_1(t)c_1^{\dagger} 
+ \sum_{l,r} b_{1lr}(t)c_1^{\dagger}a_r^{\dagger}a_l
+\sum_{l<l',r<r'} b_{1ll'rr'}(t)
c_1^{\dagger}a_r^{\dagger}a_{r'}^{\dagger}a_la_{l'}\right.
\nonumber\\
&+&\left. b_2(t)c_2^{\dagger}
+ \sum_{l,r} b_{2lr}(t)c_2^{\dagger}a_r^{\dagger}a_l
+\sum_{l<l',r<r'} b_{2ll'rr'}(t)
c_2^{\dagger}a_r^{\dagger}a_{r'}^{\dagger}a_la_{l'}+\cdots
\right ]|0\rangle,
\label{a3}
\end{eqnarray} 
where $b(t)$ are the probability amplitudes to find the 
system in the states defined by the corresponding creation and 
annihilation operators. The ``vacuum'' state $|0\rangle$ corresponds 
to the left and the right reservoirs (the emitter and the collector) 
are filled up to the Fermi levels $\mu_L$ and $\mu_R$, respectively.

Substituting Eq.~(\ref{a3}) into the Shr\"odinger equation 
$i|\dot\Psi (t)\rangle ={\cal H}|\Psi (t)\rangle$ we find an 
infinite set of equations for the amplitudes $b(t)$. Then, performing 
summation (integration) over the reservoir states 
($l,l',\ldots,r.r',\ldots$),
we can transform the Shr\"odinger equation for the amplitudes $b(t)$ into 
differential equations for the reduced density-matrix 
$\sigma^{(n)}_{i,j}(t)$ of the entire system,
where 
\begin{equation}
\sigma^{(0)}_{ij}(t)=b_i(t)b^*_j(t),~~~~
\sigma^{(1)}_{ij}(t)=\sum_{l,r}b_{ilr}(t)b^*_{jlr}(t),~~~~
\sigma^{(2)}_{ij}(t)=\sum_{ll',rr'}b_{ill'rr'}(t)b^*_{jll'rr'}(t),\; 
\cdots\,
\label{a11}
\end{equation}
and $i,j=\{ 1,2\}$ denote the occupied states of the double-dot system. 
The index $n$ denotes the number of electrons, penetrating to the right 
reservoir at time $t$. Detailed microscopic derivation of 
these equations for quantum transport can be found in\cite{gur1,gp},
and in the Appendix. 
Here we present only the final result for our system\cite{gur1}:
\begin{mathletters}
\label{c3}
\begin{eqnarray}
\dot\sigma_{11}^{(n)} & = & -D_1\sigma_{11}^{(n)}+D_1\sigma_{11}^{(n-1)}
+i\Omega_0 (\sigma_{12}^{(n)}-\sigma_{21}^{(n)})\;, 
\label{c3a}\\
\dot\sigma_{22}^{(n)} & = &-i\Omega_0 (\sigma_{12}^{(n)}-\sigma_{21}^{(n)})\;,
\label{c3b}\\
\dot\sigma_{12}^{(n)} & = & i\epsilon\sigma_{12}^{(n)}+
i\Omega_0(\sigma_{11}^{(n)}-\sigma_{22}^{(n)})
-\frac{D_1}{2}\sigma_{12}^{(n)}\, ,
\label{c3c}
\end{eqnarray}
\end{mathletters} 
where $\epsilon = E_2-E_1$, and $D_1=T_1V_d/(2\pi )$, Fig.~1.  
These equations have clear physical interpretation. Consider, for instance,
Eq.~(\ref{c3a}) for the probability rate of finding the system in the 
state, shown in Fig.~1a. The latter decays to the state 
with $(n+1)$ electrons in the collector with the rate $D_1$. 
This process is 
described by the first term in Eq.~(\ref{c3a}). On the other hand, 
there exists the opposite (``gain'') process (with the same rate $D_1$), 
when the state with $(n-1)$ in the collector 
converts to the state with $n$ electrons in the collector. 
It is described by the second term in Eq.~(\ref{c3a}). 
All these processes are generated   
by one-electron transitions between continuum states. 
If, however, one-electron transition takes place between {\em isolated} 
states, it results in a coupling between diagonal and off-diagonal 
density-matrix elements (in our case it is given by 
the last term in Eq.~(\ref{c3a})).

The evolution of the off-diagonal density-matrix elements 
$\sigma_{12}^{(n)}$ is given 
by Eq.~(\ref{c3c}). It can be interpret in the same way as the 
rate equations for the diagonal terms. 
Notice, however, the absence of the gain term in Eq.~(\ref{c3c}).
Such a term would be generated by one-electron hopping 
($n-1\to n$), resulting in $\sigma^{(n-1)}_{12}\to\sigma^{(n)}_{12}$
transition. Yet, in our case this transition is not possible, since 
the point-contact is blocked when the right dot is occupied, 
Fig.~1b.
 
Eqs.~(\ref{c3}) look as the Bloch-type optical rate equations. Yet, 
Eqs.~(\ref{c3}) were obtained from the 
many-body Schr\"odinger equation for the entire system. No stochastic 
assumptions have made in their derivation, despite the master-equations
structure of Eqs.~(\ref{c3}). In addition, these equations describe 
quantum transport on the microscopic level, in contrast with the 
usual master-equation, holding only on a coarse-grained time scale. 
 
\subsection{Time-evolution of the measured system in the presence of 
detector}

Although Eqs.~(\ref{c3}) have a rather simple form,
they describe the microscopic behavior of the measured system and 
the detector at once. In order to find the time-evolution 
of the measured system we trace out the detector 
states $n$, thus obtaining
\begin{mathletters}
\label{a6}
\begin{eqnarray}
\dot{\sigma}_{11}& = &i\Omega_0(\sigma_{12}-\sigma_{21})\;, 
\label{a6a}\\
\dot{\sigma}_{22}& = &i\Omega_0(\sigma_{21}-\sigma_{12})\;,
\label{a6b}\\
\dot{\sigma}_{12}& = & i\epsilon\sigma_{12}+i\Omega_0(\sigma_{11}
-\sigma_{22})-\frac{1}{2}\Gamma_d\sigma_{12}.
\label{a6c}
\end{eqnarray}
\end{mathletters}   
where $\sigma_{ij}=\sum_n\sigma^{(n)}_{ij}$, and 
$\Gamma_d=D_1$ is the dephasing rate generated by the detector.

As expected, the asymptotic solution 
of Eqs.~(\ref{a6}) is always the statistical mixture:
\begin{equation} 
\sigma (t)=\left (\begin{array}{cc}
\sigma_{11}(t)&\sigma_{12}(t)\\
\sigma_{21}(t)&\sigma_{22}(t)\end{array}\right )
\stackrel{\small t\to\infty}{\longrightarrow}
\left (\begin{array}{cc}
1/2&0\\0&1/2
\end{array}\right ).
\label{a8}
\end{equation}
Yet, the relevant relaxation time depends on the 
initial conditions. Consider for instance the initial conditions 
$\sigma_{11}(0)=1$,  $\sigma_{22}(0)=\sigma_{12}(0)=0$
corresponding to the electron localized in the left dot.
Solving Eqs.~(\ref{a6}) for the aligned levels ($\epsilon =0$) 
we find 
\begin{mathletters}
\label{a10}
\begin{eqnarray}
\sigma_{11}(t)&=&\frac{1}{2}
+\frac{1}{4}\left (1+\frac{\Gamma_d}{\omega}\right )
e^{-e_-t}
+\frac{1}{4}\left (1-\frac{\Gamma_d}{\omega}\right )e^{-e_+t}
\label{a10a}\\
\noalign{\vskip5pt}
\sigma_{12}(t)&=&i \frac{4\Omega_0}{\omega}\left (e^{-e_-t}
-e^{-e_+t}\right )\, ,
\label{a10b}
\end{eqnarray}
\end{mathletters}   
where $\omega =\sqrt{\Gamma_d^2-64\Omega_0^2}$, and 
$e_{\pm}={1\over4}(\Gamma_d\pm\omega)$. Therefore 
$e_+\simeq \Gamma_d/2$ and $e_-\simeq 8\Omega_0^2/\Gamma_d$
in the limit of $\Gamma_d\gg 8\Omega_0$. As a result 
the relaxation time ($\tau_Z$) {\em increases} with $\Gamma_d$: 
\begin{equation}
\tau_Z=\frac{4}{\Gamma_d-{\mbox{Re}}\ \omega}\to \frac{\Gamma_d}{8\Omega_0^2},
\;\;\; {\mbox{for}}\;\;\;  \Gamma_d\gg 8\Omega_0\, .
\label{aa8}
\end{equation}
Hence, the electron stays in the left dot for a long time, Fig.~2a, which 
is the quantum Zeno effect\cite{zeno}.
We therefore called this relaxation time as the ``Zeno'' time, $\tau_Z$.
Notice an emergence of the off-diagonal density-matrix term, 
$\sigma_{12}(t)$, that actually govern the electron transition 
between the dots during the relaxation period ($t<\tau_Z$).

If however, the electron is initially in the ground state
(the symmetric superposition), the relaxation time is much shorter
($\tau_Z\sim \Gamma_d^{-1}$), Fig.~2b. 
Indeed, in this case one  
obtains from Eq.~(\ref{a6}) that $\sigma_{11}(t)=\sigma_{22}(t)=1/2$ and 
$\sigma_{12}(t)=(1/2)\exp (-\Gamma_d t/2)$.

We thus obtained that the probability of finding the electron 
in the first dot is given by $\sigma_{11}(t)$, Eqs.~(\ref{a6}).
Consider now the point-contact current $I$. Since 
it is monitored by occupation of the first dot, Fig.~1, 
one could expect that $I(t)=I_1\sigma_{11}(t)$. 
In fact, the same result is obtained by evaluating  
of $\langle I(t)\rangle =\langle\Psi (t)|\hat I|\Psi (t)\rangle$, where 
$\hat I=i\left [{\cal H},Q_R\right ]$ and  $Q_R=e\sum_r a_r^\dagger a_r$
is the charge accumulated in the right reservoir\cite{gur1}. 
This, however, would imply 
that the detector displays the current $I_1/2$, whenever 
the electrons density-matrix becomes the mixture, Eq.~(\ref{a8}). 
On the other hand, the mixture means that the electron actually 
occupies one of the dots. As a result, the detector should 
show either $I_1$ or $0$, but not $I_1/2$. As a matter of
fact the both statements are not in a contradiction,   
since $\langle I(t)\rangle$ represents the average detector current, 
but not its actual value. Thus, the time-dependence 
of the detector current is not always determined by the electron 
density-matrix $\sigma_{11}(t)$. Actually, this cannot  
be surprising, since the observable quantity 
is the number of electrons ($n$) arriving to the collector.
The latter is given by the total density-matrix 
$\sigma_{ij}^{(n)}$, Eqs.~(\ref{c3}), which  
we are now going to evaluate. This would allow us to understand the 
detector behavior during the measurement process. 
\vskip1cm 
\begin{minipage}[t]{13cm}
\begin{center}
\leavevmode
\epsfxsize=10cm
\epsffile{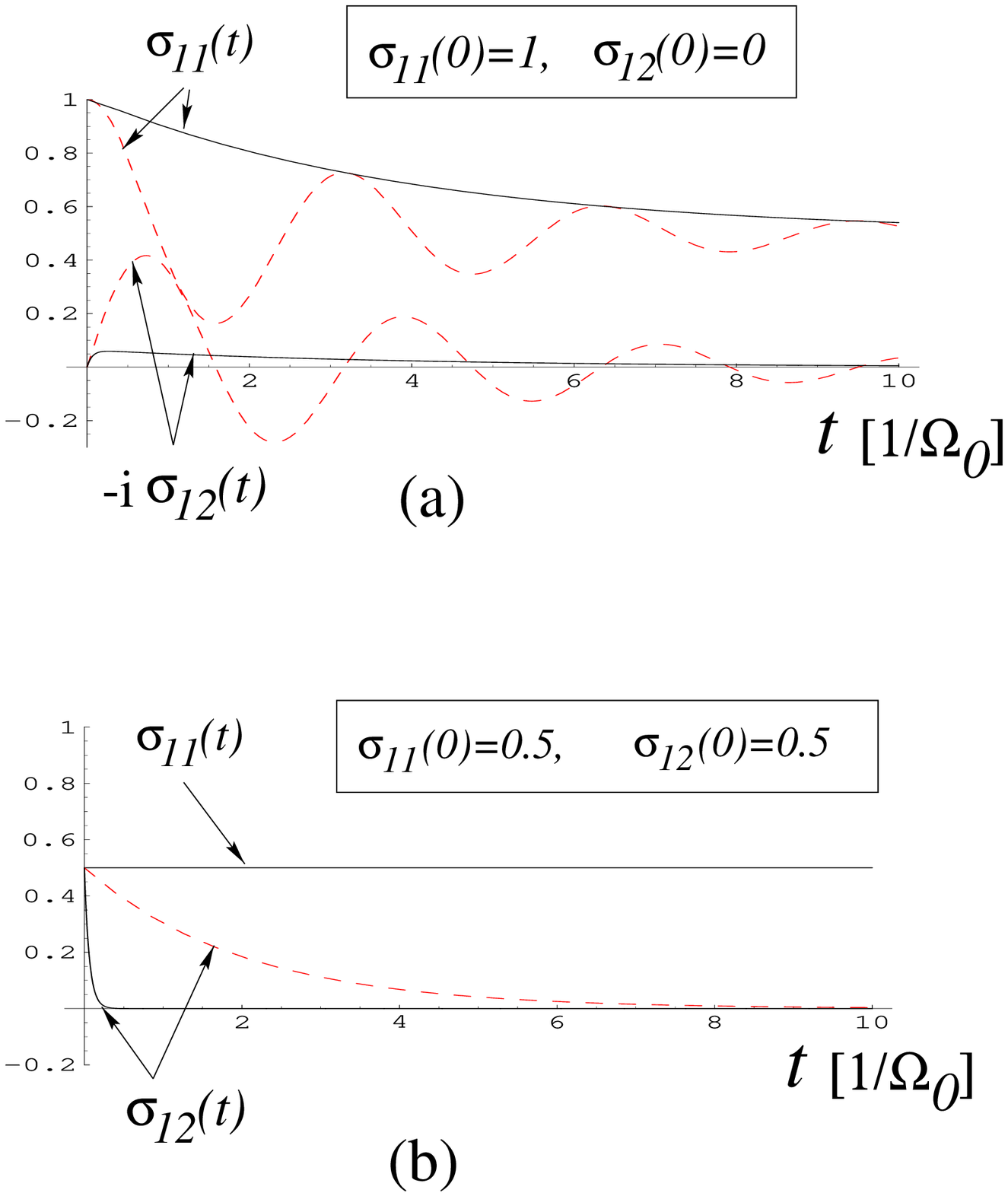}
\end{center}
{\begin{small}
Fig.~2. The occupation probability of the first well ($\sigma_{11}$) 
and the nondiagonal density-matrix element 
($\sigma_{12}$) as a function of time for $\epsilon=0$, and two
values of the dephasing rate: 
$\Gamma_d=\Omega_0$ (dashed lines), and $\Gamma_d=32\Omega_0$ (solid lines). 
The initial conditions: (a) the electron occupies the first dot,   
(b) the electron is in the ground state.
\end{small}}
\end{minipage} \\ \\  

\section{Microscopic behavior of the point-contact detector}

First we investigate the motion of carriers through 
the point-contact, decoupled from the double-dot. 
We thus put $\Omega_0=0$ in Eqs.~(\ref{c3}), so that 
the electron stays in left dot all the time.
In this case the detector is described by Eq.~(\ref{c3a}), 
which now reads
\begin{equation}
\dot p_n(t)  =  -D_1 p_n(t)+D_1 p_{n-1}(t)\ ,
\label{c1}
\end{equation}
where $p_n(t)\equiv\sigma_{11}^{(n)}(t)$ is
the probability of finding $n$ electrons in the collector at time 
$t$. The initial condition, $p_n(0)=\delta_{n0}$, 
corresponds to zero electrons in the collector. 

Eq.~(\ref{c1}) looks as a pure classical rate equation. 
Yet, no ``classical'' assumptions 
beyond quantum-mechanical treatment were made 
in its derivation\cite{gur1}. Since 
this point is very important in the following discussion, 
we present in Appendix the quantum mechanical 
derivation of Eq.~(\ref{c1}) and determine conditions for its validity.  

Eq.~(\ref{c1}) can be easily solved by applying 
the Fourier transform\cite{sasha}:    
$\tilde p(k,t)=\sum_np_n(t)\exp (ink)$.
One finds
\begin{equation}
\tilde p(k,t)=\exp\left [-D_1(1-e^{ik})t\right ] \ .  
\label{a13}
\end{equation}
The distribution $p_n(t)$ is given by the inverse Fourier transform
of $\tilde p(k,t)$. Using the stationary phase approximation we obtain
\begin{equation}
p_n(t)=\frac{1}{2\pi}\int_{-\pi}^{\pi}\tilde p(k,t)
\exp (-ink)dk\simeq \frac{1}{\sqrt{2\pi D_1t}}
\exp\left [-\frac{(D_1t-n)^2}{2D_1t}\right ].
\label{a14}
\end{equation}
This implies that $p_n(t)$ can be viewed as a wave packet of the width 
$\sqrt{2D_1t}$ propagating in the $n$-space with the 
group velocity $D_1$. Thus, the number of electrons accumulated 
in the collector is $<n(t)>=D_1t$ that corresponds to 
the detector current $I=eD_1$. Notice that $<n(t)>$ can reach 
macroscopic values in a very short time, providing that $D_1$ is large 
enough. In the following we consider the number 
of electrons accumulated in the collector as 
the observed quantity (instead of the detector current). Note that
the corresponding operator $\hat N=\sum_r a^\dagger_ra_r$ commutes 
with the operator $c^\dagger_ic_j$, so that the observation of 
$n$ does not influence the double-dot electron.    

\section{The detector behavior in the presence of double-dot}

Consider now $\Omega_0\not =0$. 
Let us evaluate the probability $P_n(t)$ of finding $n$ electrons 
in the collector at time $t$. In order to obtain this quantity we 
trace the density-matrix over 
the states of the {\em measured system} 
(cf. with Eqs.~(\ref{a6})): 
$P_n(t)=\sigma_{11}^{(n)}(t)+\sigma_{22}^{(n)}(t)$, 
where $\sigma_{ii}^{(n)}$ are given by Eqs.~(\ref{c3}).
As in the previous case we apply the Fourier transform\cite{sasha}: 
$\tilde \sigma_{ij}(k,t)=\sum_n\sigma_{ij}^{(n)}(t)\exp (ink)$. 
Then Eqs.~(\ref{c3}) become:
\begin{mathletters}
\label{c4}
\begin{eqnarray}
\dot{\tilde\sigma}_{11} & = & -D_1(1-e^{ik})\tilde\sigma_{11}
+2i\Omega_0 \Delta\tilde\sigma_{12}\;, 
\label{c4a}\\
\dot{\tilde\sigma}_{22} & = &-2i\Omega_0 \Delta\tilde\sigma_{12}\;,
\label{c4b}\\
\Delta\dot{\tilde\sigma}_{12} & = &i\Omega_0
(\tilde\sigma_{11}-\tilde\sigma_{22})-\frac{D_1}{2}
\Delta\tilde\sigma_{12}\; ,
\label{c4c}
\end{eqnarray}
\end{mathletters}
where $\Delta\tilde\sigma_{12}=
(\tilde\sigma_{12}-\tilde\sigma_{21})/2$. 
For simplicity we considered here 
the case of aligned levels, $\epsilon=0$. 
Note that $D_1\equiv\Gamma_d$ is the decoherence rate 
in Eq.~(\ref{a6c}). Next, by applying the Laplace transform, 
$s(k,E)=\int_0^\infty\tilde\sigma (k,t)\exp (iEt)dt$, 
we reduce Eqs.~(\ref{c4}) to a system of linear algebraic equations:
\begin{equation}
\left(\begin{array}{ccc}E+iD_1(1-e^{ik})&0&
2\Omega_0\\
0&E&-2\Omega_0\\
\Omega_0&-\Omega_0&E+iD_1/2\end{array}\right )
\left(\begin{array}{c}s_{11}\\s_{22}\\ \Delta s_{12}
\end{array}\right )=
\left ( \begin{array}{c}i\sigma_{11}^{(0)}(0)\\
i\sigma_{22}^{(0)}(0)\\ 
{\mbox {Im}}\ \sigma_{21}^{(0)}(0)
\end{array}\right )
\label{c5}
\end{equation}
where the r.h.s. is defined by the initial condition. 

Solving Eqs.~(\ref{c5}) and performing the inverse Laplace and
Fourier transformations we obtain
\begin{equation}
P_n(t)=\sum_{\scriptstyle j=1\atop
\scriptstyle (j'>j''\not =j)}^3
\int_{-\pi}^{\pi}{dk\over 2\pi} 
{{\cal M}(e_j) \over
(e_j-e_{j'})(e_j-e_{j''})}\exp (-ie_jt-ink)\, ,
\label{a18}
\end{equation}
where $e_{1,2,3}$ are the roots of the secular 
determinant and ${\cal M}$ is the corresponding
minor determinant. 
The secular determinant is represented    
by a cubic equation. We solve it perturbatively 
in the limits of weak  and strong decoherence (damping). 

\subsection{Weak damping ($D_1\ll \Omega_0$)}

Consider for the definiteness
the initial conditions 
$\sigma_{11}^{(0)}(0)=1$ and 
$\sigma_{22}^{(0)}(0)= \sigma_{12}^{(0)}(0)=0$, 
corresponding to the electron localized in the left dot,
Then ${\cal M}=e_j(e_j+iD_1/2)-4\Omega_0^2$ in Eq.~(\ref{a18}).
The roots of the secular equation in the case of weak damping 
are $e_1\simeq -iD_1\xi /2$ and 
$e_{2,3}\simeq \pm 2\Omega_0$, where $\xi=1-\exp (ik)$.
It is clear from Eq.~(\ref{a18}) 
that the dominant contribution is coming from the first root. Using 
the stationary phase approximation we find the following expression for 
$P_n(t)$
\begin{equation}
P_n(t)=\frac{1}{\sqrt{\pi D_1t}}
\exp\left [-\frac{\left 
(D_1t/2-n\right )^2}{D_1t}\right ]. 
\label{a20}
\end{equation}
It looks as Eq.~(\ref{a14}) for the undistorted motion of carriers 
through the point-contact, except for  the  group velocity, $D_1/2$. 
The latter corresponds to the ``average'' value of the detector current 
($I_1/2$).  The interpretation of this result is very simple. Since 
$\Omega_0\gg D_1$, the observed electron oscillates many times 
between the dots during the time of an electron penetration 
to the collector. As a result the detector current displays 
the average electron charge ($e/2$) in each of the dots. 

In fact, the same average current $I_1/2$ would be displayed
even for $D_1\simeq \Omega_0$.  
This can be seen from Fig.~3 that shows $P_n(t)$, obtained from  
a numerical solution of Eqs.~(\ref{c3}) for $D_1=\Omega_0$ (bars), 
in a comparison with the distribution $p_n(t)$
(solid lines), obtained from
Eq.~(\ref{c1}) for $D_1=\Omega_0/2$. It follows from this figure
that the distribution of electrons arriving to the collector 
corresponds to the current $I_1/2$ flowing through the point-contact. 
This implies that the electron oscillations between the dots, shown in 
Fig.~2a by the dashed line would not be reflected 
in the behavior of the point-contact current. 
Therefore, the point-contact cannot be a good 
detector in the case of weak damping\cite{kor}. 

This example clearly demonstrates that the time-dependence 
of the detector current $I_d(t)$ is not fully determined by 
the density-matrix of the observed electron via the relation  
$I_d(t)=eD_1\sigma_{11}(t)$\cite{gur1}. The latter 
represents an ensemble average over the electron in the double-dot.
In each particular experiment, however, the electron oscillation 
between the dots cannot be seen. We believe that this is 
the reason of a disagreement with recent predictions for the 
detector current behavior for the case of weak damping\cite{stod,hacken}.  

\vskip0.2cm 
\begin{minipage}{13cm}
\begin{center}
\leavevmode
\epsfxsize=13cm
\epsffile{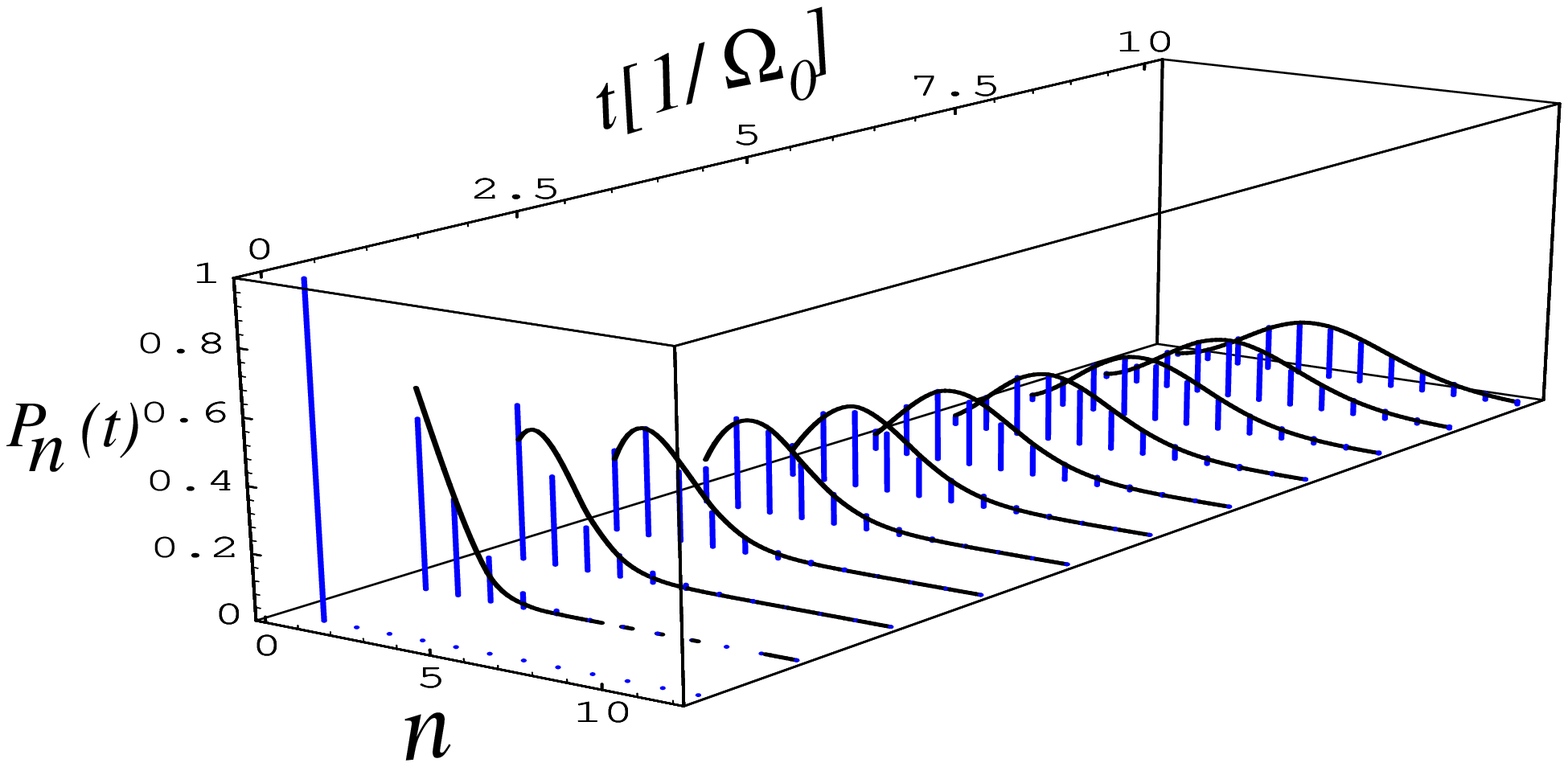}
\end{center}
{\begin{small}
Fig.~3. Probability distribution of electrons in the collector,
$P_n(t)$, for $D_1=\Omega_0$.
The electron is initially localized in the left dot.  
The solid lines represent smooth interpolation 
of $p_n(t)$, Eq.~(\ref{c1}) for $D_1=\Omega_0/2$.
\end{small}}
\end{minipage} \\ \\ 

\subsection{Strong damping ($D_1\gg \Omega_0$)}

\subsubsection{The electron is initially in the left dot}

Consider now the strong 
decoherence limit, $D_1\gg\Omega_0$. In this case
many electrons can penetrate to the collector during one oscillation 
of the observed electron. Thus, the point-contact should 
represent a good detector in this case. However, the 
strong decoherence would result in quantum Zeno effect, 
which hinders the electron transitions between the dots. 
Let us investigate how this effect is reflected in the 
detector behavior. 

Solving the secular equation 
perturbatively, we obtain the following expressions for 
the roots $e_j$, Eq.~(\ref{a18})
\begin{mathletters}
\label{a19}
\begin{eqnarray}
e_{1,2}&=&-i\left ({D_1\xi\over 2}+{4\Omega_0^2\over 
D_1}\mp\sqrt{{D_1^2\xi^2\over 4}+{16\Omega_0^4\over D_1^2}}
\right )+{\cal O}\left [{\Omega_0^2\over D_1}\right ]^2\, ,
\label{a19a}\\
\noalign{\vskip5pt}
e_3&=&-i{D_1\over 2}+i{8\Omega_0^2\over D_1}(1+\xi)
+{\cal O}\left [{\Omega_0^2\over D_1}\right ]^2\, .
\label{a19b}
\end{eqnarray}
\end{mathletters}
where $\xi=1-\exp (ik)$. The minor determinant
${\cal M}$ in Eq.~(\ref{a18}) is the same as in 
the previous case.  Substituting 
Eqs.~(\ref{a19}) into Eq.~(\ref{a18}) we find that the 
main contribution is coming from the roots $e_{1,2}$.
The contribution from the third root is
$\propto\exp (-D_1t/2)$, and therefore it can be  
important only at very short times, $t<2/D_1$. 

Consider first the distribution $P_n(t)$ in the time-interval  
$2/D_1\lesssim t\lesssim \tau_Z$, when the electron stays localized 
in the first dot (Fig.~2a). Then 
the main contribution to the integral (\ref{a18}) is coming 
from $k\simeq 1/n\gg 8\Omega^2_0/D_1^2$. In this region
$e_1\simeq -4i\Omega_0^2/D_1$ 
and $e_2\simeq -iD_1\xi+e_1$, Eq.~(\ref{a19a}).  
Substituting these values 
into Eq.~(\ref{a18}) and neglecting the terms of higher orders 
in $\Omega_0^2/(D_1^2\xi)$ and $\xi$ we find that 
$P_n(t)=p_n(t)$, Eq.~(\ref{a14}). This means that the distribution 
$P_n(t)$ does indeed correspond to the electron localized in the left dot. 
Thus, the detector would display the current $I_1=eD_1$ during 
the time-interval $t\lesssim \tau_Z$, in an agreement with 
the behavior of the reduced electron density-matrix,
Eq.~(\ref{a10}), Fig.~2a.

In order to confirm the validity of the above result  
we show in Fig.~4 the distribution of $P_n(t)$ (bars),
obtained from the numerical solution of Eqs.~(\ref{c3}) for 
$0\le t\le \Omega_0^{-1}$ and $D_1=32\Omega_0$. 
For comparison we display the distribution $p_n(t)$,
Eq.~(\ref{c1}), corresponding to the current $I_1=eD_1$ 
flowing through the contact. 
One finds that that in an agreement with our analytical calculations. 
the distributions $P_n(t)$ and $p_n$ are practically indistinguishable.
 
Consider now the probability distribution $P_n(t)$ for 
$t\gg \tau_Z$, when the 
electron density matrix becomes the mixture, Eq.~(\ref{a8}).
It corresponds to $k\simeq 1/n \ll 8\Omega_0^2/D_1^2$ in 
Eq.~(\ref{a18}). Using Eq.~(\ref{a19a}) one finds that 
$e_1\simeq -iD_1\xi/2$ and $e_2\simeq -8i\Omega_0^2/D_1+e_1$. 
This implies that the term $\propto\exp (-ie_2t)$ in Eq.~(\ref{a18}) 
is exponentially suppressed for $t\gg \tau_Z$. Eventually, it is only the 
term $\propto\exp (-ie_1t)$ that survives in the ``asymptotic'' 
limit. As a result we arrive to Eq.~(\ref{a20}) for 
$P_n(t)$ in the limit of $t\gg \tau_Z$. This represents the ``average'' 
detector current ($I_1/2$). Actually, one can easily demonstrate  
that the ``asymptotic'' behavior of  $P_n(t)$,
given by Eq.~(\ref{a20}) is valid for any relation between 
$D_1$ and $\Omega_0$, and for any initial conditions. 
\vskip0.2cm 
\begin{minipage}{13cm}
\begin{center}
\leavevmode
\epsfxsize=13cm
\epsffile{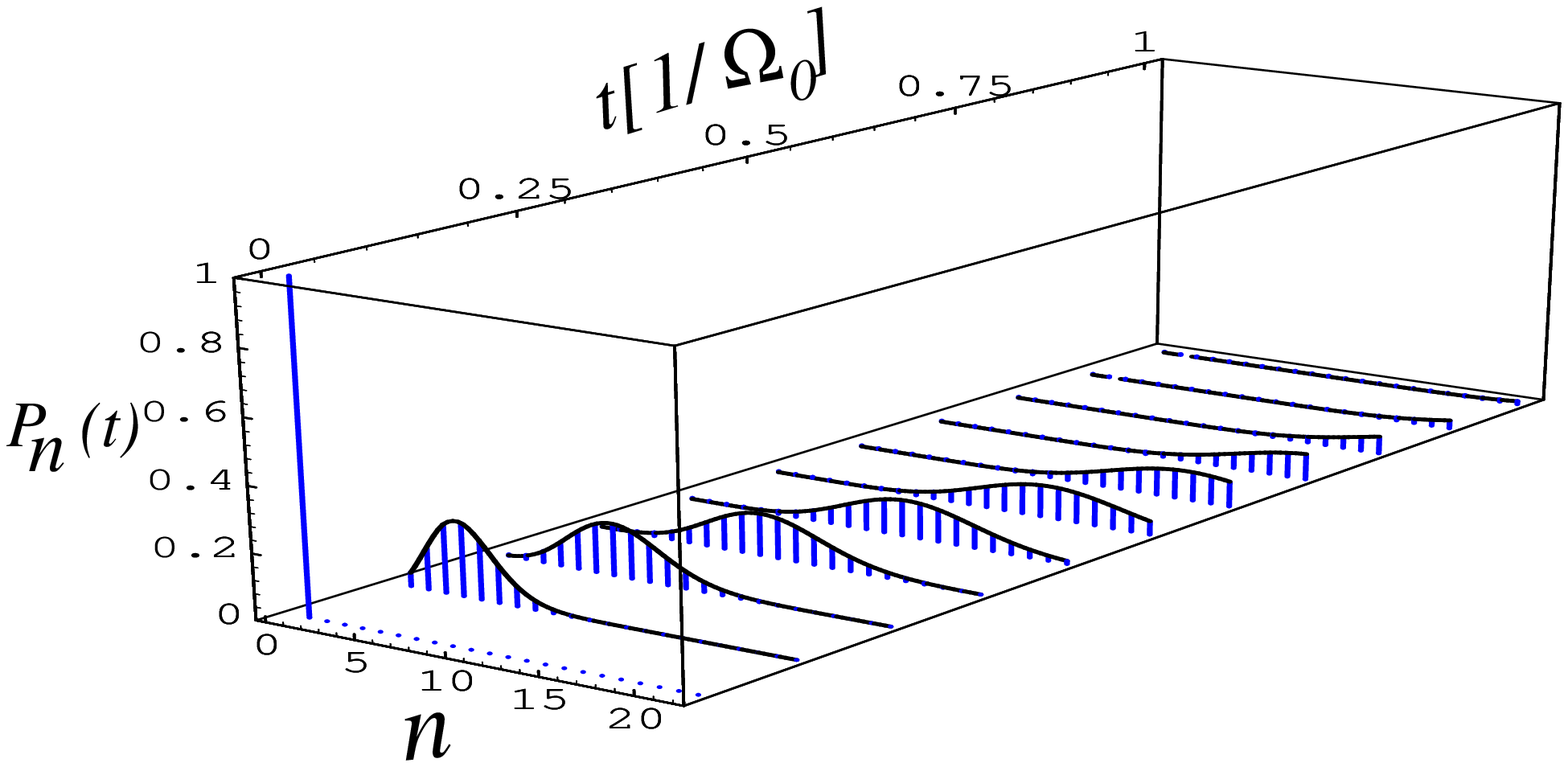}
\end{center}
{\begin{small}
Fig.~4. Probability distribution of electrons in the collector,
$P_n(t)$, for $t\le\Omega_0^{-1}$ and $D_1=32\Omega_0$.
The electron is initially localized in the left dot.  
The solid lines represent smooth interpolation 
of $p_n(t)$, Eq.~(\ref{c1}), corresponding to the electron permanently 
localized in the left dot. 
\end{small}}
\end{minipage} \\ \\ 

Thus, we found that in the case of strong damping and the 
electron is initially localized in one of the dots,  
the detector current behaves in an accordance with 
the electron density-matrix, $I_d(t)=eD_1\sigma_{11}(t)$.  
Indeed, the detector displays the 
current $I_1$ for $t\lesssim \tau_Z$, and the average current, $I_1/2$, 
for $t\gg \tau_Z$, i.e. when 
the electron density-matrix becomes the mixture. 

\subsubsection{The electron is initially in the ground state}

Let us solve Eq.~(\ref{c5}) for the initial conditions 
$\sigma_{11}^{(0)}(0)=\sigma_{22}^{(0)}(0)=
\sigma_{12}^{(0)}(0)={1\over2}$,
corresponding to the double-dot electron in the ground state.
(Actually, the same result would be obtained if the electron is initially 
in the statistical mixture). Then 
\begin{equation}
{\cal M}= e_j[2e_j+iD_1(1+\xi) ]-{1\over2}D_1^2\xi -8\Omega_0^2
\label{a21}
\end{equation}
in Eq.~(\ref{a18}). 
If the detector behavior is determined by the 
electron density-matrix $\sigma_{ij}(t)$, than  
the distribution $P_n(t)$ should display the ``average'' current  
$I_1/2$, corresponding to Eq.~(\ref{a20}), already for $t\gg 2/D_1$,
when $\sigma_{ij}(t)$ becomes the the mixture, (Fig.~2b).
Yet, it is not the case. One obtains from 
Eq.~(\ref{a18}) 
\begin{equation}
P_n(t)={1\over2}\delta_{n,0}
\exp \left (-{4\Omega_0^2\over D_1}t\right )
+{1\over2}p_n(t)\, ,
\label{a22}
\end{equation}
where $p_n(t)$ is given by Eq.~(\ref{a14}), and we neglected  
the terms of higher orders in   
$\Omega_0^2/(D_1^2\xi)$ and $\xi$. Thus we find that
$P_n(t)$,  given by Eq.~(\ref{a22}), is very 
different from that given by Eq.~(\ref{a20}), despite of
the corresponding electron density matrix 
is almost the statistical mixture in the both cases.

In order to confirm our analytical calculations we present 
in Fig.~5 the distribution $P_n(t)$ (bars) as a function 
of $n$ and $t$, found from the numerical solution of 
Eqs.~(\ref{c3}) for $0\le t\le \Omega_0^{-1}$ and 
$D_1=32\Omega_0$. The solid lines represent ${1\over2}p_n(t)$
as given by Eq.~(\ref{c1}). Thus, Eq.~(\ref{a22}) represents 
the exact result quite well. 
It is also in a qualitative agreement with the recent numerical 
calculations\cite{sasha}.

It follows from Eq.~(\ref{a22}) and Fig.~5 that  
$P_n(t)$ displays two  
peaks, at $n=0$ and $n=D_1t$, which remain separated 
for a long time ($\sim \tau_Z$). This implies that 
one finds either zero or $n\simeq D_1t$ electrons 
with the probability $1/2$ at any time 
$1/D_1\lesssim t \lesssim \tau_Z$. If we assume that an 
actual observation of the accumulated charge $(en)$ does not 
affect the evolution of $P_n(t)$, then the repeated observations 
would show unlimited fluctuations of electron charge in the collector 
(for $\tau_Z=D_1/8\Omega_0^2\to\infty$).
Such a scenario looks impossible. Hence, we have to assume 
that the ``actual fact'' happens at some time, so that  
one of these two possibilities is actually realized. 
At that moment the evolution of $P_n(t)$ starts with the new initial 
conditions. It looks as a quantum jump\cite{knight}, 
which however, is not generated 
by quantum evolution of the entire system, Eqs.~(\ref{c3}). 
Now we consider this problem in more details. 
\vskip0.2cm 
\begin{minipage}{13cm}
\begin{center}
\leavevmode
\epsfxsize=13cm
\epsffile{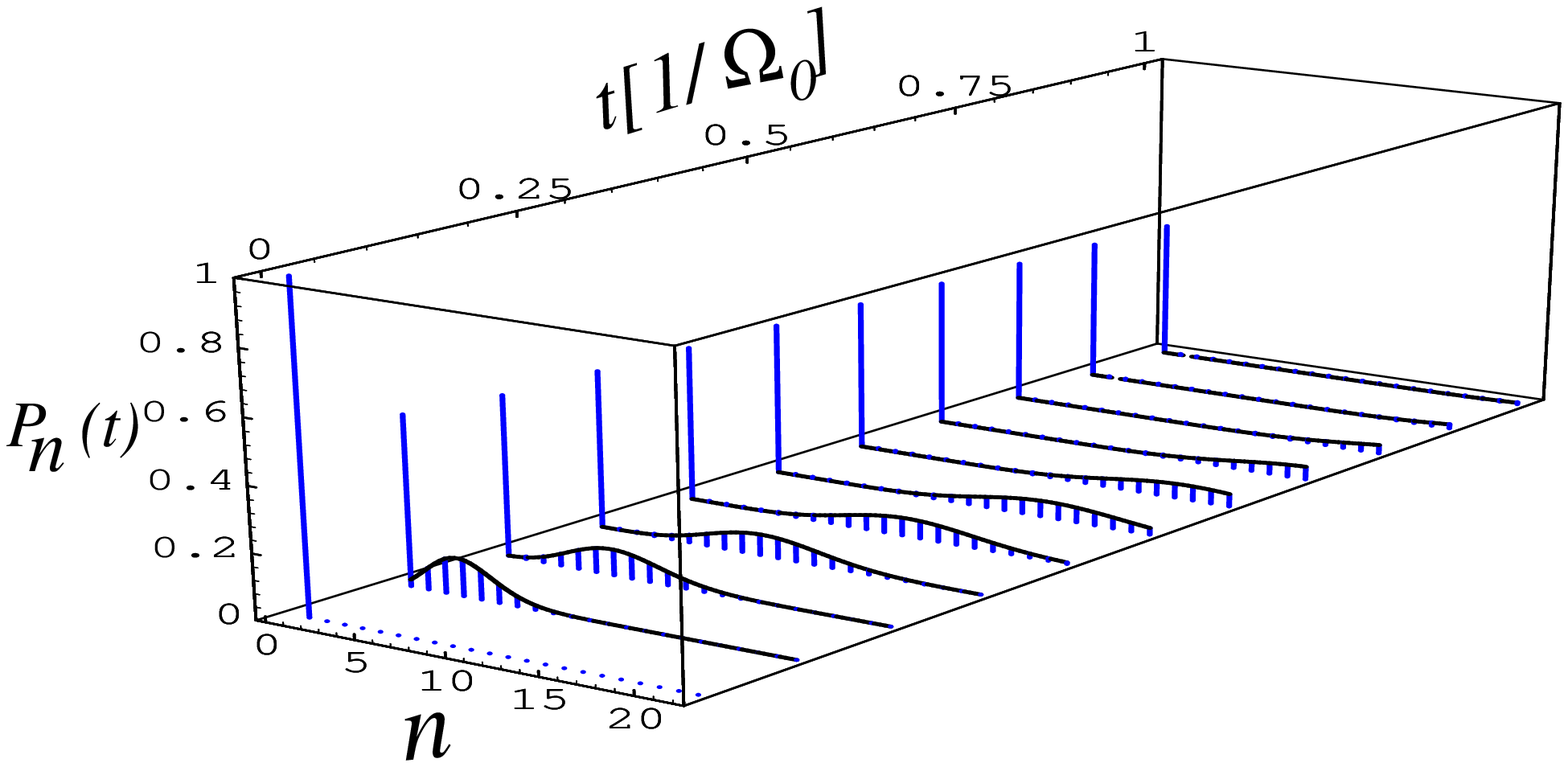}
\end{center}
{\begin{small}
Fig.~5. Probability distribution of 
$P_n(t)$ for $t\le\Omega_0^{-1}$ for the initial condition 
corresponding to the electron in the symmetric superposition,
and $D_1=32\Omega_0$. The solid lines show the smooth interpolation 
of ${1\over2}p_n(t)$, Eq.~(\ref{c1}). 
\end{small}}
\end{minipage} \\ \\ 

\section{Collapse and the role of observer}

We demonstrated in the previous example that an 
uncertainty of finding the electron 
in one of the dots leads to the possibility of observing 
either zero or large number of electrons in the collector. 
Such a macroscopic amplification of the 
single electron quantum state 
resembles the Schr\"odinger cat paradox. Similarly, it is rather 
natural to assume that the actual realization of the 
electron position inside the double-dot takes place  
before the observer looks on the detector. 
Yet, it still must be related to 
the measurement process that detects 
the occupied dot. However, it is not clear what does constitute
the detection in our case. For instance, 
one can suggest that a penetration of {\em one} 
electron to the collector is already the measurement, since it 
indicates that the left dot is occupied.
On the other hand, such a single electron is not actually observed.
Thus, one can define a measurement in an alternative way,  
as a process that outcome  
is stored in a stable form, ready to be retrieved by 
an observer\cite{home}. In this case the right 
reservoir of the detector must be coupled with an another macroscopic 
device (the ``pointer'') that actually counts the accumulated charge. 
If such a pointer responds only to a macroscopic amount of charge   
accumulated in the collector, the measurement 
would be completed much later than in the first case. 

\subsection{Continuous collapse}

Let us begin with the first scenario. In this case the ``measurement 
time''\cite{sasha}, $t_{ms}=1/D_1$, is the time of one-electron  
penetration to the collector when  
the left dot is occupied. If we assume that each time-interval $t_{ms}$
the system starts its evolution with the new initial conditions,  
even if the collector charge is not actually counted,
we arrive to a continuous collapse model (c.f.\cite{pearle}).   
Yet, such an extreme scenario would result in significant violation 
of the quantum mechanical predictions. Indeed, consider the initial 
conditions corresponding to the blocked point-contact 
(the right dot is occupied) and the right reservoir is empty:  
$\sigma_{22}^{(n)}(0)=\delta_{0n}$ and 
$\sigma_{11}^{(n)}(0)=\sigma_{12}^{(n)}(0)=0$ in Eqs.~(\ref{c3}).
One easily obtains from Eqs.~(\ref{c3}) that the probability of 
finding zero electrons in the collector at $t=t_{ms}$ is
$P_0(t_{ms})\simeq 1-2\Omega_0^2t_{ms}^2$.
Then after $D_1t$ such successive measurements the probability of finding 
the system undecayed at the time $t$ is
\begin{equation}
P_0(t)=\left (1-{2\Omega_0^2\over D_1^2}\right )^{D_1t}
+{\cal O}\left ({\Omega_0\over D_1}\right )^4\simeq 
1-{2\Omega_0^2\over D_1} t
\label{z1}
\end{equation}
Therefore the point-contact remains blocked (Zeno time) 
for the time-interval $D_1/2\Omega_0^2$. On the other hand, 
the non-interrupted evolution of Eqs.~(\ref{c3}) yields 
four times less value ($D_1/8\Omega_0^2$)
for the corresponding Zeno time, Eq.~(\ref{aa8}).
Thus the continues collapse scenario predicts 
different dwell-time for the observed electron, in a 
comparison with the Schr\"odinger evolution. 
Notice, however, that if we choose     
the measurement time as corresponding to penetration 
of four electrons, $t_{ms}=4/D_1$, the continuous collapse 
would yield the same average dwell-time, as the Schr\"odinger evolution
(c.f.\cite{schul}). Yet, even in this case, the distribution $P_0(t)$ 
as a function of time would be still different.

In fact, the predictions of the continuous collapse model 
and the quantum mechanics  
can be checked experimentally by switch on the pointer 
at a corresponding time. 
However, it is quite clear that the continuous collapse
means {\em strong} violation of the quantum mechanics. 
In order to avoid it, we assert  that whenever 
the ``actual fact'' is realized, the entire system proceeds 
its quantum evolution with no interruption for a long (Zeno) time $\tau_Z$, 
Eq.~(\ref{aa8}). This time is  completely defined
by the Schr\"odinger equation.
Yet, the main question still remains: when these mysterious quantum 
jumps can take a place. We claim here that this problem can also 
be investigated experimentally. Such a unique possibility 
is provided by the large Zeno time $\tau_Z$ that can reach 
macroscopic time-scales for $D_1/\Omega_0\to\infty$.
  
\subsection {Spontaneous collapse}

Consider again the double-dot electron, which is initially 
in the ground state (symmetric superposition). As we  
demonstrated above that the Schr\"odinger 
evolution of the entire system generates two peaks in 
the distribution $P_n(t)$, Fig.~5, corresponding to  
the electron localized in the right or in the left dot. 
Let us assume that one of these possibilities is always 
realized on the microscopic time-scale, 
i.e. for $t\sim 1/D_1$. 
Then the electron would stay in the same dot for a long time, $\tau_Z$.
Afterwards it can be found in the second dot 
with the probability $1/2$.
Hence, one can expect that the number of electrons 
accumulated in the collector, $N(t)$, would 
display the following behavior, as shown schematically
in Fig.~6 by the solid line. 

In fact, there is no a-priory time-scale for these jumps. 
For instance, a localization of the observed electron 
in one of the dots can happen on the time-scale 
$t_0\gg 1/D_1$, when the number of electrons in the collector
($D_1t_0$) becomes macroscopically large. 
In order to determine $t_0$ experimentally 
we connect the right reservoir with a macroscopic 
``pointer'' that actually counts electrons in the 
collector, and displays the relevant data  
directly to the observer. We assume 
that the pointer starts the counting only after 
the number of electrons in the collector reaches some 
threshold value $\bar N$, i.e. for $t>\bar t=\bar N/D_1$. 
This threshold $\bar N$ can be varied by the experimentalist 
in a wide interval. But we always assume that $\bar t<\tau_Z$.   
\vskip0.4cm 
\begin{minipage}{13cm}
\begin{center}
\leavevmode
\epsfxsize=11cm
\epsffile{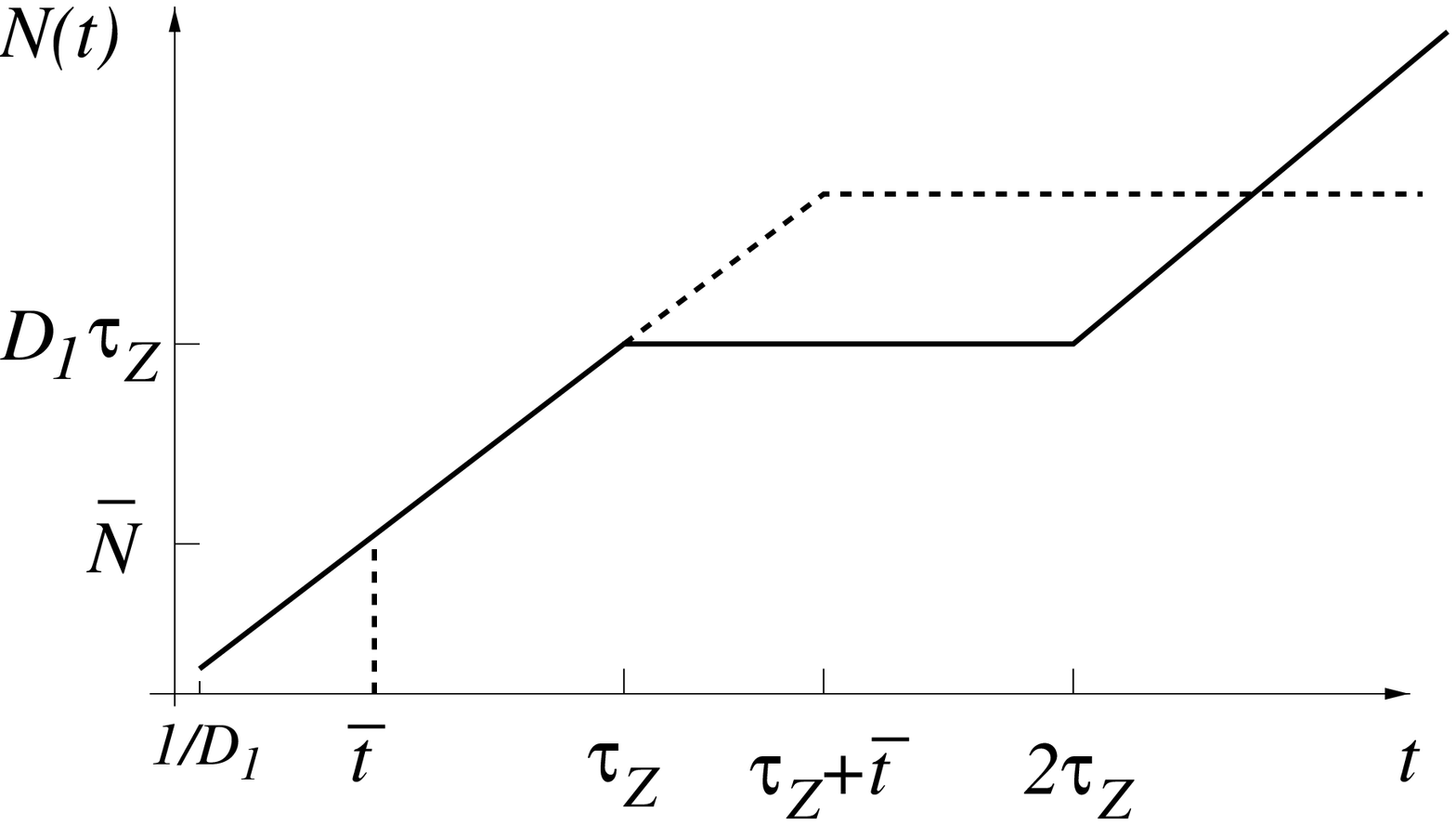}
\end{center}
{\begin{small}
Fig.~6. The number of electron penetrating to the collector
as a function of time. The dashed line corresponds to the actual
pointer display if the quantum jump happens at $\bar t=\bar N/D_1$,
where $\bar N$ is the pointer threshold. 
\end{small}}
\end{minipage} \\ \\ 

The entire system can be described quantum-mechanically 
by adding the corresponding terms in the total 
Hamiltonian, Eq.~(\ref{a1})
\begin{equation}
{\cal H}={\cal H}_{PC}+{\cal H}_{DD}+{\cal H}_{P}+{\cal H}_{int}
+{\cal H}'_{int},
\label{p1}
\end{equation}
where ${\cal H}_{P}$ and ${\cal H}'_{int}$ describe the pointer 
and its interaction with electrons in the collector. These terms 
can be written in a form of the tunneling Hamiltonian, Eqs.~(\ref{a2}), 
where the threshold  $\bar N$ 
can be accounted for in ${\cal H}'_{int}$ by increasing the number of 
the corresponding creation and annihilation operators.
Note that the pointer does not interact with the double-dot. 
It is rather clear that such it cannot 
influence the time-evolution of $P_n(t)$, except for a 
modification of the decoherence rate $D_1\to D'_1$.
The latter in fact, can be made arbitrary small, 
$\Delta D_1/D_1\to 0$.  

Let us assume that the pointer threshold time 
$ \bar t> t_0$. It implies that the pointer displays 
the electron localization in the left dot at $t=\bar t$, 
i.e. later than it actually happened. As a result, the electron 
dwell-time shown by the pointer becomes 
$\tau_Z+t_0-\bar t<\tau_Z$. 
Therefore, by decreasing the pointer threshold $\bar N$,
the dwell-time shown by the pointer would always increase,  
unless $\bar t < t_0$. It would allow us to determine 
the time-scale for the spontaneous collapse.

\subsection{Collapse due to observation}

The above scenarios of spontaneous collapse 
assume the existence of parameters that determine 
the collapse time-scale. However, these 
parameters cannot be obtained within 
the Quantum Mechanics. It would imply 
that the Quantum Mechanics is not 
a complete theory. On the other hand, no such 
parameters are needed if we assume that the collapse takes 
place whenever the relevant information on 
the system can be directly available 
to the observer. Although this point of view  
looks very strange, it is rather close to  
the von-Neumann and even to the ``orthodox'' (Copenhagen) interpretation 
of the measurement in quantum mechanics\cite{neu}. 
Most important, however, that such a scenario has definite 
experimental consequences in the framework of our setup. 
Indeed, in this case the collapse should {\em always} take place
whenever the pointer starts to respond to an accumulated charge,  
i.e. at the time $\bar t$. This is because the left dot is 
{\em definitely} occupied when the pointer displays the charge. Otherwise,
the point-contact is blocked and the collector charge 
cannot reach its threshold value $\bar N$.
(The same arguments are applied, when 
the pointer does not show any collector charge at $t=\bar t$). 
As a result, the pointer would display the same dwell-time $\tau_Z$  
(the dashed line in Fig.~6) for any values of $\bar N$. 

Let us demonstrate on a simple example 
that such a scenario is somehow inherent
in the quantum-mechanical approach. 
Consider the point-contact detector detached 
from the double-dot. In this case the detector behavior is 
described by Eq.~(\ref{c1}) that
gives the probability distribution $p_n(t)$, Eq.~(\ref{a14}), 
of finding $n$ electrons in the collector at time $t$. 
Although this equation was obtained from the many-body 
Schr\"odinger equation (see Appendix), it can be viewed as the 
classical probabilistic equation. However, the latter 
implies that its outcome does 
depend on the information, obtained by an observer 
at any intermediate stages (the Bayes principle).
Indeed, let us assume that the detector displays $N_1$ electrons 
at some time $t_1$. Then it is quite clear that 
the distribution $p_n(t)$ for $t\ge t_1$ is given by the same  
Eq.~(\ref{c1}), but with new initial conditions 
at $t=t_1$\cite{ozawa}. One obtains
\begin{equation}
p_n(t)\simeq \frac{1}{\sqrt{2\pi D_1(t-t_1)}}
\exp\left [-\frac{(D_1t-n+\Delta N)^2}{2D_1(t-t_1)}\right ]\, ,
\label{a23}
\end{equation}
where $\Delta N=N_1-D_1t_1$. Comparing this 
result with Eq.~(\ref{a14}) we find that 
$p_n(t)$ has the same group velocity 
(the average current), but the width of the distribution is narrower. 
That means that the information acquired during the measurement,    
affects the electron probability distribution $p_n(t)$.
(Actually, it affects the fluctuation, but not the average value
of the observed quantity). Notice that this result is quite natural
from the classical point of view. Indeed, the probabilistic description 
of classical systems is not a complete one. The measurement just 
improves our knowledge on the system, so the statistical 
uncertainty diminishes. 

As a matter of fact, Eq.~(\ref{c1}) has been derived
quantum-mechanically, from the many-body Schr\"odinger equation. 
Nevertheless, we must obtain  the same Eq.~(\ref{a23}) for $p_n(t)$, 
if the number of electrons in the collector 
is actually known at $t=t_1$. 
Otherwise, the quantum mechanics does
not reproduce its classical limit. It means that whenever 
the pointer starts to deliver the relevant information to an observer
(i.e. at $t_1=\bar t$), the probability distribution should jump 
from Eq.~(\ref{a14}) to Eq.~(\ref{a23}).
The same should be expected when the detector is coupled 
with the double-dot. This would imply that 
the moment of quantum jumps is determined by 
the pointer sensitivity. 

\section{Discussion}

It is well-know that the quantum mechanics gives only 
a {\em conditional} probability. The same holds for the 
classical probabilistic description. 
Therefore an information on the system obtained 
by a measurement, changes the probabilities 
of the following events (the wave-function collapse). 
In the classical case this obviously cannot influence 
the true system behavior, providing that system is not 
distorted by the detector. Indeed, the probabilities 
in the classical case represent only our ignorance as to 
the true state. 

The situation with the quantum mechanical probabilities 
is not so clear. The probability amplitudes there must be ascribe 
some objective meaning independent of human knowledge\cite{jaynes}.
For instance, an information on the system true state 
implies the vanishing of the corresponding off-diagonal 
density-matrix elements, which can influence the system behavior.
Such an effect cannot be totally attributed due to the decoherence, as
can be seen in an example of continuous measurement. 
In this case the apparatus (and also  
the environment) generates decoherence, which steady 
diminishes the off-diagonal density matrix elements, but 
never makes them {\em zero}. It manifests itself in the quantum Zeno effect 
which slows down quantum transitions between the isolated states.
Nevertheless these transitions, which proceed via  
the off-diagonal density-matrix elements, are 
never totally interrupted by the decoherence.
The same remains valid by extending the quantum mechanical 
description via von Neumann hierarchy\cite{neu} (a system ``measured''
by another system etc). In this way one can only increase 
the decoherence rate. Yet, an ``observation'' which nullifies 
the off-diagonal density-matrix elements would not be singled out.   

This severe problem of quantum measurement cannot be resolved by 
using only the theoretical arguments.  
In this paper we proposed to investigate    
the wave-function collapse in experiments with  
continuous monitoring of a single quantum system.
Such experiments are now within reach of present technology.  
The main idea consists in 
freezing the system in the same state for a very long time 
(the Zeno time, $\tau_Z$) due to its continuous monitoring.
The Zeno time is evaluated by using the Schr\"odinger equation for 
the entire system, including the detector, connected with 
a macroscopic pointer. The latter is switched on automatically
at $t=t_0$, where $t_0<\tau_Z$. (The ``switching on'' can be made 
gradually, but on the scale much smaller than $\tau_Z$). 
Then, if the collapse takes place before the system 
is observed, the calculated Zeno 
time would be different from that shown by the pointer. 
Since $t_0$ varies with the pointer threshold, 
it would be possible to single out experimentally the 
time-scale for the wave-function collapse. 
This would be extremely important 
in understanding the nature of the collapse and whether it is  
related with the actual information available to the observer. 

We have not investigated in this paper all possible consequences of 
the measurement collapse. For instance, the measurement with 
weakly responding detector\cite{kor}, or the influence of 
AC voltage applied across the point-contact detector\cite{hacken}.
Yet, we believe that our model for continuous measurement of a 
single system is suitable for investigation of difference
sides of the measurement problem and allows their experimental
realization. We also expect that this model can be adapted
for optical experiments with a single atom\cite{knight}. 
  
\section{Acknowledgments}
I owe special thanks to E. Buks for attracting my attention 
to the problem of detector current and numerous fruitful 
discussions. Special thanks to Y. Aharonov for very useful  
discussions, which helped me in elaboration of different aspects 
of the measurement problem. I am also grateful to M. Heiblum, M. Kleber, 
A.  Korotkov, M. Marinov, L. Pitaevskii and D. Sprinzak, 
for useful discussions. 

\appendix 

\section{Quantum-mechanical derivation of rate equations 
for a point-contact detector}

Here we present the derivation of classical  
rate equations (\ref{c1}), starting from the Schr\"odinger equation.
By using the same technique we also 
derive the average current and the current 
fluctuations and compare our results with those existing in the 
literature. 

Consider the point-contact detector, describing by 
the tunneling Hamiltonian,
${\cal H}_{PC}$, Eq.~(\ref{a2a}). The initial 
(vacuum) state ($|0\rangle$) is the state, where the levels in 
the emitter and the collector are initially filled up to the Fermi 
energies $\mu_L$ and $\mu_R$ respectively, Fig.~1.
The many-body wave function for this system can be written 
in the occupation number representation as 
\begin{equation}
|\Psi (t)\rangle = \left [ b_0(t) + \sum_{l,r} b_{lr}(t)a_r^{\dagger}a_l
+\sum_{l<l',r<r'} b_{ll'rr'}(t)a_r^{\dagger}a_{r'}^{\dagger}a_la_{l'}
+\cdots\right ]|0\rangle\ , 
\label{ap2}
\end{equation} 
where $b(t)$ are the time-dependent probability amplitudes to
find the system in the corresponding states with the initial 
condition $b_0(0)=1$, and all the other $b(0)$'s being zeros
(cf. with Eqs.~(\ref{a3}) for the entire system). 
Substituting it into the Shr\"odinger equation 
$i|\dot\Psi (t)\rangle ={\cal H}_{PC}|\Psi (t)\rangle$ 
and performing the Laplace transform:  
$\tilde{b}(E)=\int_0^{\infty}e^{iEt}b(t)dt$
we obtain an infinite set of the coupled equations for the 
amplitudes $\tilde b(E)$:
\begin{mathletters}
\label{ap4}
\begin{eqnarray}
& &E \tilde{b}_{0}(E) - \sum_{l,r} \Omega_{lr}\tilde{b}_{lr}(E)=i
\label{ap4a}\\
&(&E + E_{l} - E_r) \tilde{b}_{lr}(E) - \Omega_{lr}\tilde{b}_0(E) 
-\sum_{l',r'}\Omega_{l'r'}\tilde{b}_{ll'rr'}(E)=0
\label{ap4b}\\
&(&E + E_{l}+E_{l'} - E_r-E_{r'}) \tilde{b}_{ll'rr'}(E) 
- \Omega_{l'r'}\tilde{b}_{lr}(E)+\Omega_{lr}\tilde{b}_{l'r'}(E)
\nonumber\\
&&~~~~~~~~~~~~~~~~~~~~~~~~~~~~~~~~~~~~~~~~~~~~~~~~~~~~~
-\sum_{l'',r''}\Omega_{l''r''}\tilde{b}_{ll'l''rr'r''}(E)=0
\label{ap4c}\\
& &\cdots\cdots\cdots\cdots\cdots 
\nonumber
\end{eqnarray}
\end{mathletters}

Eqs. (\ref{ap4}) can be substantially simplified by replacing 
the amplitude $\tilde b$ in 
the term $\sum\Omega\tilde b$ of each of the equations  by 
its expression obtained from the subsequent equation\cite{gur1,gp}.  
For example,   
substituting $\tilde{b}_{lr}(E)$ from Eq.~(\ref{ap4b}) into 
Eq.~(\ref{ap4a}), 
one obtains
\begin{equation}
\left [ E - \sum_{l,r}\frac{\Omega^2}{E + E_{l} - E_r}
    \right ] \tilde{b}_{0}(E) - \sum_{ll',rr'}
    \frac{\Omega^2}{E + E_{l} - E_r}\tilde{b}_{ll'rr'}(E)=i,
\label{ap5}
\end{equation}
where we assumed that the hopping amplitudes 
are weakly dependent functions on the energies
$\Omega_{lr}\equiv\Omega (E_l,E_r)=\Omega$.
Since the states in the reservoirs are very dense (continuum), 
one can replace the sums over $l$ and $r$ by integrals, for instance  
$\sum_{l,r}\;\rightarrow\;\int \rho_{L}(E_{l})\rho_{R}(E_{r})
\,dE_{l}dE_r\:$,
where $\rho_{L,R}$ are the density of states in the emitter and collector. 
Then the first sum in Eq.~(\ref{ap5}) becomes an
integral which can be split into a sum of the singular and principal value 
parts. The singular part yields $i\pi\Omega^2\rho_L\rho_R V_d$,
and the principal part is merely included into
redefinition of the energy levels. The second sum 
(non-factorized term) in Eq.~({\ref{ap5}) 
can be neglected in the limit of large bias $V_d\gg \Omega^2\rho$. 
Indeed, by replacing 
$\tilde{b}_{ll'rr'}(E)\equiv \tilde{b} (E,E_l,E_{l'},E_r,E_{r'})$ and 
the sums by the integrals we find that the integrand   
has the poles in $E_{l,r}$-variables on the same sides 
of the integration contours. It means that 
the corresponding integral vanishes.

Applying analogous considerations to the other equations of the
system (\ref{ap4}), we finally arrive to the following set of equations: 
\begin{mathletters}
\label{ap6}
\begin{eqnarray}
&& (E + iD/2) \tilde{b}_{0}=i
\label{ap6a}\\
&& (E + E_{l} - E_r + iD/2) \tilde{b}_{lr}
      - \Omega\tilde{b}_{0}=0
\label{ap6b}\\ 
&& (E + E_{l}+ E_{l'} - E_{r} - E_{r'} + iD/2) \tilde{b}_{ll'rr'} -
      \Omega\tilde{b}_{lr}+\Omega \tilde{b}_{l'r'}=0,
\label{ap6c}\\
& &\cdots\cdots\cdots\cdots\cdots 
\nonumber
\end{eqnarray}
\end{mathletters}
where $D=2\pi\Omega^2\rho_L\rho_R V_d$. 

The amplitudes $\tilde b$ are directly related with the corresponding 
probabilities, $p_n(t)$, of finding $n$ electrons in the collector:
\begin{equation}
p_0(t)=|b_0(t)|^2,~~~~p_1(t)=\sum_{l,r}|b_{lr}(t)|^2,~~~~
p_2(t)=\sum_{ll',rr'}|b_{ll'rr'}(t)|^2,\; \cdots\ .
\label{ap8}
\end{equation}  
By applying the inverse Laplace transform 
\begin{equation}
p_n(t)=
\sum_{l\ldots , r\ldots}
\int \frac{dEdE'}{4\pi^2}\tilde b_{l\cdots r\cdots}(E)
\tilde b^*_{l\cdots r\cdots}(E')e^{i(E'-E)t}\, .
\label{ap9}
\end{equation}
one can transform  Eqs.~(\ref{ap6})
into the rate equations for $p_n(t)$ (c.f.\cite{gur1,gp}). We find 
\begin{mathletters}
\label{ap10}
\begin{eqnarray}  
&&\dot p_0(t) = -Dp_0(t)
\label{ap10a}\\
&&\dot p_1(t) = Dp_0(t)-Dp_1(t)
\label{ap10b}\\
&&\dot p_2(t) = Dp_1(t)-Dp_2(t)
\label{ap10c}\\
&&\cdots\cdots\cdots\cdots\cdots 
\nonumber
\end{eqnarray}
\end{mathletters}
which are the classical rate equations (\ref{c1}).

The essential point in our quantum-mechanical derivation 
of Eqs.~(\ref{ap10}) is neglect of the ``cross'' terms,
namely those where the amplitudes $\tilde b$ cannot be 
factorized out (like the second term in Eq.~(\ref{ap5})). 
As a result we obtain Eqs.~(\ref{ap6}), which lead eventually to the 
rate equations (\ref{ap10}). Although the neglect of the cross-terms 
can be justified in the limit of $V_d\gg \Omega^2\rho$, 
we nevertheless expect that Eqs.~(\ref{ap6}) 
are valid even beyond that limit. 
For instance, we demonstrate below, that these equations result 
in correct expressions for the average current and the current 
fluctuations. 

\subsection{Average current}

The current operator is defined as a commutator of the accumulated charge 
with the Hamiltonian
\begin{equation}
\hat I=ie\left [{\cal H}_{PC},\sum_r a_r^\dagger a_r\right ]=
ie\sum_{l,r}\Omega_{lr}(a^\dagger_la_r-a^\dagger_ra_l)
\label{ap11}
\end{equation}
Using Eqs.~(\ref{ap2}), (\ref{ap11}) we find the following 
expression for the average current
\begin{equation} 
I(t) =\langle\Psi (t)|\hat I|\Psi (t)\rangle=
2 e\ {\mbox {Im}}\left [\sum_{l,r}\Omega_{lr}
\ b_0(t)b^*_{lr}(t)+\sum_{ll'rr'}
\Omega_{l'r'}\ b_{lr}(t)b^*_{ll'rr'}(t)+\cdots\right ]
\label{ap12}
\end{equation}
As in the previous consideration we replace the sums by 
the integrals. By applying the inverse Laplace transform and using 
Eqs.~(\ref{ap6}) we can carry out all the integrations analytically. 
For instance, 
\begin{equation} 
\sum_{l,r}\Omega_{lr}b^*_{lr}(t)=\int {dE\over 2\pi}e^{iEt}\int 
{\rho_L\rho_R\Omega^2\ \tilde b_0^*(E)\over E+E_L-E_R}dE_LdE_R=
i\pi\rho_L\rho_R eV_d\ \Omega^2\ b_0^*(t)
\label{ap13}
\end{equation}

The same procedure can be applied for other terms in Eq.~(\ref{ap12}).
Finally, by taken into account the normalization condition 
of the wave function,  $\langle\Psi (t)|\Psi (t)\rangle=1$, one finds
\begin{equation}
I(t) =2 e^2\pi\rho_L\rho_R\Omega^2V_d\left [
|b_0(t)|^2+\sum_{l,r}|b_{lr}(t)|^2+\cdots\right]=2e^2\pi
\rho_L\rho_R\Omega^2V_d\, .
\label{ap14}
\end{equation} 
By using $(2\pi )^2\Omega^2\rho_L\rho_R=T$\cite{bard}, 
where $T$ is the transmission probability, we can rewrite 
the current as $I=e^2 T\, V_d/(2\pi )$. 
This coincides with the well known Landauer formula\cite{land}.

\subsection{Current fluctuations}

Let us evaluate the average of $\hat{I}^2$ operator. 
Using Eq.~(\ref{ap11}) one can write 
\begin{equation}
\langle\Psi (t)|{\hat I}^2|\Psi (t)\rangle=
-e^2\langle\Psi (t)|\sum_{lr,l'r'}\Omega_{lr}\Omega_{l'r'}
(a^\dagger_la_r-a^\dagger_ra_l)
(a^\dagger_{l'}a_{r'}-a^\dagger_{r'}a_{l'})|\Psi (t)\rangle\, .
\label{ap15}
\end{equation} 
Now we split this sum into two parts corresponding to 
$lr\not =l'r'$ and $lr=l'r'$. Then, using Eq.~(\ref{ap2}) 
we rewrite Eq.~(\ref{ap15}) as 
\begin{equation}
\langle\Psi (t)|{\hat I}^2|\Psi (t)\rangle=
4 e^2\ {\mbox{Re}}\left [\sum_{lr,l'r'}\Omega_{lr}\Omega_{l'r'}
b_0(t)b^*_{ll'rr'}(t)+\cdots\right ]+2 e^2\sum_{lr}\Omega_{lr}^2\, .
\label{ap16}
\end{equation} 
The second term can be written as $2e\Delta\nu I$, where 
$\Delta \nu$ is the 
band width and $I$ is the average current, Eq.~(\ref{ap14}). 
This expression corresponds to Shottky noise. Consider now 
the first sum in Eq.~(\ref{ap16}). It can be evaluated in 
the same way as Eq.~(\ref{ap12}). Indeed, by replacing 
the sum by the integral and using the inverse Laplace 
transform and Eqs.~(\ref{ap6}) we obtain
\begin{equation} 
\sum_{lr,l'r'}\Omega_{lr}\Omega_{l'r'}b^*_{ll'rr'}(t)
=\int {dE\over 2\pi}e^{iEt}\int 
{\rho^2_L\rho^2_R\Omega^4\ \tilde b_0^*(E)dE_LdE'_LdE_RdE'_R
\over(E+E_L+E'_L-E_R-E'_R)(E+E_L-E_R)}
\label{ap17}
\end{equation} 
Taking into account the contributions from the poles and the normalization 
of the wave function we find that the first term of Eq.~(\ref{ap16}) can 
be represented as $-e^3\ T^2 V_d\Delta \nu/2\pi$. Finally the average 
current fluctuation can be written as 
\begin{equation}
\langle (\Delta I)^2\rangle =2 e\ \Delta\nu {e^2V_d\over 2\pi}T(1-T)\, .
\label{ap18}
\end{equation} 
This coincides with the result obtained earlier 
by using different techniques\cite{but1}.

\end{document}